\documentclass[prl,aps,showpacs,twocolumn]{revtex4}
\usepackage{graphicx}

\begin{document}
\title{Origin of nonlinear contribution to the shift of the critical
temperature in atomic Bose-Einstein condensates}
\author{Sergei Sergeenkov$^{1}$, Fabio Briscese$^{1,2}$, Marcela Grether$^{3}
$ and M. de Llano$^{4}$}
\affiliation{$^{1}$Departamento de F\'{\i}sica, CCEN, Universidade Federal da Para\'{\i}%
ba, Cidade Universit\'{a}ria, 58051-970 Jo\~{a}o Pessoa, PB, Brazil\\
$^{2}$Istituto Nazionale di Alta Matematica Francesco Severi,
Gruppo Nazionale di Fisica Matematica, 00185 Rome, EU\\
$^{3}$Facultad de Ciencias, Universidad Nacional Aut\'{o}noma de
M\'{e}xico,
04510 M\'{e}xico, DF, M\'{e}xico\\
$^{4}$Instituto de Investigaciones en Materiales, Universidad Nacional Aut%
\'{o}noma de M\'{e}xico, A. P. 70-360, 04510 M\'{e}xico, DF,
M\'{e}xico}


\date{\today}

\begin{abstract}
We discuss a possible origin of the experimentally
observed nonlinear contribution to the shift $\Delta
T_{c}=T_c-T_{c}^{0}$ of the critical temperature $T_{c}$ in an
atomic Bose-Einstein condensate (BEC) with respect to the critical temperature $%
T_{c}^{0}$ of an ideal gas. We found that accounting for a
nonlinear (quadratic) Zeeman effect (with applied magnetic field
closely matching a Feshbach resonance field $B_0$) in the
mean-field approximation results in a rather significant
renormalization of the field-free nonlinear contribution $b_{2}$,
namely $\Delta T_{c}/T_{c}^{0}\simeq b_{2}^{\ast }(a/\lambda
_{T})^{2}$ (where
$a$ is the s-wave scattering length, $\lambda _{T}$ is the thermal wavelength at $T_{c}^{0}$) with $%
b_{2}^{\ast }=\gamma ^{2}b_{2}$ and $\gamma =\gamma (B_0)$. In
particular, we predict $b_{2}^{\ast }\simeq 42.3$ for the
$B_{0}\simeq 403G$ resonance observed in the $\ ^{39}K$ BEC.
\end{abstract}

\pacs{67.85.Hj, 67.85.Jk}

\maketitle

Studies of Bose-Einstein condensates (BECs) continue to be an important
subject in modern physics (see, e.g., Refs.\cite%
{dalfovo,ps,parkinswalls,burnett} and further references therein).
Atomic BECs are produced in the laboratory in laser-cooled,
magnetically-trapped
ultra-cold bosonic clouds of different atomic species (including $%
^{87}Rb$ \cite{Ander,Cornish},$^{7}Li$ \cite{Bradley}, $%
^{23}Na$ \cite{Davis}, $^{1}H$ \cite{Fried}, $^{4}He$ \cite%
{Pereira}, $^{41}K$ \cite{Mondugno}, $^{133}Cs$ \cite{Grimm}, $%
^{174}{Yb}$ \cite{Takasu03} and $^{52}Cr$ \cite{Griesmaier}, among
others). Also, a discussion of a relativistic BEC has appeared in Ref.\cite%
{PRL07Baker} and BECs of photons are most recently under investigation \cite%
{becs of photons}. In addition, BECs are successfully utilized in cosmology
and astrophysics \cite{becs cosmology} as they have been shown to constrain
quantum gravity models \cite{briscese}.

In the context of atomic BECs interparticle interactions must play a
fundamental role since they are necessary to drive the atomic cloud to
thermal equilibrium. Thus, they must be carefully taken into account when
studying the properties of the condensate. For instance, interatomic
interactions change the condensation temperature $T_{c}$ of a BEC, as was
pointed out first by Lee and Yang \cite{a1,a2} (see also Refs.\cite%
{a3,a4,a5,a6,a7,a8,a9,a10,a11,a12} for more recent works).

The first studies of interactions effects were focused on
\textit{uniform} BECs. Here, interactions are irrelevant in the
mean field (MF) approximation (see Refs.\cite{a7,a10,a11,a12}) but
they produce a shift in the condensation temperature of uniform
BECs with respect to the ideal noninteracting case, which is due
to quantum correlations between bosons near the critical point.
This effect has been finally quantified in \cite{a7,a8} as $\Delta
T_{c}/T_{c}^{0}\simeq 1.8(a/\lambda _{T})$, where $\Delta
T_{c}\equiv T_{c}-T_{c}^{0}$ with $T_{c}$ the critical temperature
of the gas of interacting bosons, $T_{c}^{0}$ is the BEC
condensation temperature in the ideal noninteracting case, $a$ is
the s-wave scattering length used to represent interparticle
interactions \cite{dalfovo,parkinswalls,burnett}, and $\lambda
_{T}\equiv \sqrt{2\pi \hbar ^{2}/m_{a}k_{B}T_{c}^{0}}$ is the
thermal wavelength for temperature $T_{c}^{0}$ with $m_{a}$ the
atomic mass.

But laboratory condensates are not uniform BECs since they are
produced in atomic clouds confined in magnetic traps. For trapped
BECs, interactions affect the condensation temperature even in the
MF approximation, and the shift in $T_{c}$ in terms of the s-wave
scattering length $a$ is given by
\begin{equation}
\Delta T_{c}/T_{c}^{0}\simeq b_{1}(a/\lambda _{T})+b_{2}(a/\lambda _{T})^{2}.
\label{deltaTsuNONTUNIFORM1}
\end{equation}%
with $b_{1}\simeq -3.4$ \cite{dalfovo} and $b_{2}\simeq 18.8$ \cite{briscese
EPJB}.

High precision measurements \cite{condensatePRL} of the
condensation temperature of $^{39}K$ in the range of parameters
$N\simeq (2-8)\times 10^{5}$,  $10^{-3}<a/\lambda _{T}<6\times
10^{-2}$ and $T_{c}\simeq (180-330)nK$ have detected second-order
(nonlinear) effects in $\Delta T_{c}/T_{c}^{0}$ fitted by the
expression $\Delta T_{c}/T_{c}^{0}=b_{2}^{exp}(a/\lambda
_{T})^{2}$ with $b_{2}^{exp}\simeq 46\pm 5$. This result has been
achieved exploiting the high-field $403G$
Feshbach resonance in the $|F,m_{F}>=|1,1>$ hyperfine (HF) state of a $%
^{39}K $ condensate where $F\equiv S+I$ is the total spin of the atom with $S
$ and $I$ being electron and nuclear spin, respectively, and $m_{F}$ is the
projection quantum number. Thus, the theoretically predicted \cite{briscese
EPJB} quadratic-amplitude coefficient $b_{2}$ turned out to be in a rather
strong disagreement with the available experimental data. There have also
been some efforts to theoretically estimate the correct value of $b_{2}$ in
the MF approximation by considering anharmonic and even
temperature-dependent traps \cite{castellanos}, which however have not been
too successful. Therefore one could expect that a more realistic prediction
of the experimental value of $b_{2}^{exp}$ should take into account some
other so far unaccounted effects.

The main goal of this paper is to show that, taking into account
the nonlinear (quadratic) Zeeman effect and using the MF
approximation, it is quite possible to explain the experimentally
observed \cite{condensatePRL} value of $b_2$ for the $403G$
resonance of the hyperfine $|F,m_F> = |1,1>$ state of $^{39}K$
with no need to go beyond-MF approximation.

Recall that experimentally the s-wave scattering length parameter
$a$ is tuned via the Feshbach-resonance technique based on Zeeman
splitting of bosonic atom levels in an applied magnetic field.
This means that the interaction constant $g\equiv (4\pi \hbar
^{2}a/m_{a})$ is actually \textit{always} field-dependent. More
explicitly, according to the interpretation of the Feshbach
resonance \cite{derrico,williams}

\begin{equation}  \label{aB}
a(B)=a_{bg}\left( 1-\frac{\Delta }{B-B_{0}}\right)
\end{equation}
where $a_{bg}$ is a so-called background value of $a$, $B_{0}$ is
the resonance peak field, and $\Delta $ the width of the
resonance.

Thus, in order to properly address the problem of
condensation-temperature shifts (which are always observed under
application of a nonzero magnetic field $B$), one must account for
a Zeeman-like contribution. It should be emphasized, however, that
a single (free) atom Zeeman effect (induced by either electronic
or nuclear spin) $\mu _{a}B$ is not important for the problem at
hand simply because it can be accounted for by an appropriate
modification of the chemical potential.

Recall that in the presence of a linear Zeeman effect, the basic
properties of an atomic BEC can be understood within the so-called
"condensate wave function" approximation \cite{ps}
\begin{equation}
{\cal H} =\int d^3x H(x)
\end{equation}
where $H(x)=gn^2-E_Zn$ with $n(x)=\Psi^{+}(x)\Psi(x)$ being the
local density of the condensate ($\Psi(x)$ is the properly defined
wave function of macroscopic condensate), and $E_Z=\mu _BB$ the
Zeeman energy (with $\mu_B$ being the Bohr magneton).

Following Bogoliubov's recipe \cite{Bogoliubov}, let us consider
small deviation of the condensate fraction from the ground state
$n_0$ (the number) by assuming that $n(x)\simeq n_0+\delta n(x)$
with $\delta n(x)\ll n_0$. Treating, as usually, $n_0$ and $\delta
n(x)$ independently, we obtain from Eq.(3) that in the presence of
the linear Zeeman effect $gn_0=E_Z$ (meaning that $E_Z$ is playing
a role of the chemical potential \cite{nonlinear1}) and, as a
result, the BEC favors the following energy minimum:
\begin{equation}
\delta H_0(x)=2gn_0\delta n(x)-E_Z\delta n(x)=gn_0\delta n(x)
\end{equation}

Thus, we come to the conclusion that at low magnetic fields (where
the linear Zeeman effect is valid), in accordance with the
available experimental results \cite{nonlinear1}, there is no any
tangible change of the BEC properties (including $g$
modification). On the other hand, there is a clear-cut
experimental evidence \cite{nonlinear2,nonlinear3} in favor of the
so-called Breit-Rabi nonlinear (quadratic) HF-mediated Zeeman
effect \cite{breit} in BEC. We are going to demonstrate now how
this nonlinear phenomenon (which is not a trivial generalization
of the linear Zeeman effect) affects the BEC properties (including
a feasible condensation temperature shift). Recall that in strong
magnetic fields, the magnetic-field energy shift of the sublevel
$m_F$ of an alkali-metal-atom ground state can be approximated
(with a rather good accuracy) by the following expression
\cite{nonlinear2}

\begin{equation}  \label{HF Hamiltonian}
E_{NLZ} = A_{HF}\frac{E_Z^2}{h\delta \nu_{hf}}
\end{equation}
where $A_{HF}=\left[1-\frac{4m_F^2}{(2I+1)^2}\right]$, and $\delta
\nu_{hf}$ is the so-called hyperfine splitting frequency between
two ground states.

Now, by repeating the above-mentioned Bogoliubov's procedure, we
obtain a rather nontrivial result for BEC modification. Namely, it
can be easily verified that HF-mediated nonlinear Zeeman effect
gives rise to the following two {\it equivalent} options for the
energy minimization (based on the previously defined ground state
with $E_Z=gn_0$): (a) $E_{NLZ}\propto g^2n_0^2$ or (b)
$E_{NLZ}\propto E_Zgn_0=(\mu _BB) gn_0$. As a matter of fact, the
choice between these two options is quite simple. We have to
choose (b) simply because (a) introduces the second order
interaction effects ($\propto g^2$) which are neglected in the
initial Hamiltonian (3). As a result, the high-field nonlinear
Zeeman effect produces the following modification of the local BEC
energy:
\begin{equation}
\delta H_{NLZ}(x) \simeq 2gn_0\delta
n(x)+A_{HF}gn_0\left(\frac{\mu_BB}{h\delta \nu_{hf}}\right)\delta
n(x)
\end{equation}

Therefore, accounting for nonlinear Zeeman contribution will
directly  result in a renormalization of the high-field scattering
length
\begin{equation}
a^{\ast }=a\left(1+\frac{1}{2}A_{HF}\frac{\mu_BB}{h\delta
\nu_{hf}}\right)
\end{equation}

Now, by inverting (\ref{aB}) and expanding the resulting $B(a)$
dependence into the Taylor series (under the experimentally
satisfied conditions $a_{bg}\ll a$ and $\Delta \ll B_{0}$)
\begin{equation}
B(a)\simeq B_0\left \{1-\frac{\Delta}{B_0}\left[
\left(\frac{a_{bg}}{a}\right)+\left(\frac{a_{bg}}{a}\right)^2+...\right]
\right\}
\end{equation}
one obtains
\begin{equation}  \label{a B 2}
a^{\ast }\simeq \gamma a+O(a_{bg}/a, \Delta /B_{0})
\end{equation}%
for an explicit form of the renormalized scattering length due to
Breit-Rabi-Zeeman splitting with
\begin{equation}
\gamma \equiv 1+\frac{1}{2}A_{HF}\left(\frac{\mu_BB_0}{h\delta
\nu_{hf}}\right)
\end{equation}

To find the change in $b_{2}$ in the presence of the quadratic
Zeeman effect one simply replaces the original (Zeeman-free) scattering length $a$ in (\ref%
{deltaTsuNONTUNIFORM1}) with its renormalized form $a^{\ast }$ given by (%
\ref{a B 2}), which results in a nonlinear contribution to the shift of the
critical temperature, specifically%
\begin{equation}  \label{delta2}
\frac{\Delta T_{c}}{T_{c}^{0}}\simeq b_{2}\left( \frac{a^{\ast
}}{\lambda _{T}}\right) ^{2}
\end{equation}%
Furthermore, by using (\ref{a B 2}), one can rewrite (\ref%
{delta2}) in terms of the original scattering length $a$ and
renormalized amplitude $b_{2}^{\ast }$ as follows%
\begin{equation}  \label{delta3}
\frac{\Delta T_{c}}{T_{c}^{0}}\simeq b_{2}^{\ast }\left(
\frac{a}{\lambda _{T}}\right) ^{2}
\end{equation}%
where the coefficient due to the Breit-Rabi-Zeeman contribution is%
\begin{equation}
b_{2}^{\ast }\simeq \gamma ^{2}b_{2}  \label{b star 3}
\end{equation}%
with $\gamma$ defined earlier.

Let us consider the particular case of the $B_{0}\simeq 403G$
resonance of the hyperfine $|F,m_{F}>=|1,1>$ state of $^{39}K$.
For this case \cite{new},  $S=1/2$, $m_{F}=1$, $I=3/2$, and
$\delta \nu_{hf}\simeq 468MHz$. These parameters produce
$A_{HF}=3/4$ and $\gamma \simeq 1.5$ which readily leads to the
following estimate of the quadratic amplitude contribution due to
the HF mediated Breit-Rabi-Zeeman effect, $b_{2}^{\ast }\simeq
2.25b_{2}\simeq 42.3$ (using the mean-field value $b_{2}\simeq
18.8$ \cite{briscese EPJB}), in a good agreement with the
observations \cite{condensatePRL}. It is interesting to point out
that the obtained value of $\gamma$ for $^{39}K$ BEC is a result
of a practically perfect match between the two participating
energies: Zeeman contribution at the Feshbach resonance field,
$\mu_BB_0\simeq 4\times 10^{-25}J$, and the contribution due to
Breit-Rabi hyperfine splitting between two ground states, $h\delta
\nu_{hf}\simeq 3\times 10^{-25}J$.

And finally, an important comment is in order regarding the
applicability of the present approach (based on the Taylor
expansion of (2)) to the field-induced modification of the linear
contribution (defined via the amplitude $b_{1}$ in (1)) to the
shift in $T_{c}$. According to the experimental curve depicting
$\Delta T_{c}$ vs $a/\lambda _{T}$ behavior, the linear
contribution is limited by $10^{-3}\leq a/\lambda _{T}\leq 5\times
10^{-3}$. Within the Feshbach-resonance interpretation, this
corresponds to a low-field ratio $a/a_{bg}\simeq 1$ which
invalidates the Taylor expansion scenario based on using a small parameter $%
a_{bg}/a\ll 1$ applicable in high fields only. Besides, as we have
demonstrated earlier, the linear Zeeman effect (valid at low
fields only) is not responsible for any tangible changes of BEC
properties. Therefore, another approach is needed to properly
address the field-induced variation (if any) of the linear
contribution $b_1$.

To conclude, it was shown that accounting for a
hyperfine-interaction induced Breit-Rabi nonlinear (quadratic)
Zeeman term in the mean-field approximation can explain the
experimentally observed shift in the critical temperature $T_{c}$ for the $%
^{39}K$ condensate. It would be interesting to subject the
predicted universal relation (13) to a further experimental test
to verify whether or not it can also explain the shift in other
bosonic-atom condensates.

This work was financially supported by the Brazilian agencies CNPq and
CAPES. MdeLl thanks PAPIIT-UNAM for grant IN-100314 and MG for grant
IN-116914, both Mexico.

\end{document}